\documentclass[conference]{IEEEtran}
\IEEEoverridecommandlockouts
\usepackage{cite}
\usepackage{amsmath,amssymb,amsfonts}
\usepackage{algorithm}
\usepackage{algorithmicx}
\usepackage[noend]{algpseudocode}

\usepackage{graphicx}
\usepackage{textcomp}
 \usepackage{array,multirow,graphicx}
 \usepackage{float}
 \usepackage{url}

\usepackage{mypackage}

\def\BibTeX{{\rm B\kern-.05em{\sc i\kern-.025em b}\kern-.08em
    T\kern-.1667em\lower.7ex\hbox{E}\kern-.125emX}}
\begin{document}

\title{Interpreted Execution of Business Process Models on Blockchain}

\author{\IEEEauthorblockN{1\textsuperscript{st} Orlenys~L{\'{o}}pez{-}Pintado}
\IEEEauthorblockA{\textit{University of Tartu} \\
Tartu, Estonia \\
orlenyslp@ut.ee}
\and
\IEEEauthorblockN{2\textsuperscript{nd} Marlon Dumas}
\IEEEauthorblockA{\textit{University of Tartu} \\
Tartu, Estonia \\
marlon.dumas@ut.ee}
\and
\IEEEauthorblockN{3\textsuperscript{rd} Luciano Garc{\'{i}}a{-}Ba{\~{n}}uelos}
\IEEEauthorblockA{\textit{Tecnol{\'{o}}gico de Monterrey} \\
Monterrey, Mexico \\
luciano.garcia@tec.mx}
\and
\IEEEauthorblockN{4\textsuperscript{th} Ingo Weber}
\IEEEauthorblockA{\textit{Data61, CSIRO} \\
Sydney, Australia \\
ingo.weber@data61.csiro.au}
}

\maketitle

\graphicspath{{figs/}}

\begin{abstract}
Blockchain technology provides a tamper-proof mechanism to execute inter-organizational business processes involving mutually untrusted parties. Existing approaches to blockchain-based process execution are based on code generation. In these approaches, a process model is compiled into one or more smart contracts, which are then deployed on a blockchain platform. 
Given the immutability of the deployed smart contracts, these compiled approaches ensure that all process instances conform to the process model.
However, this advantage comes at the price of inflexibility. Any changes to the process model require the redeployment of the smart contracts (a costly operation). In addition, changes cannot be applied to running process instances.
To address this lack of flexibility, this paper presents an interpreter of BPMN process models based on dynamic data structures. 
The proposed interpreter is embedded in a business process execution system with a modular multi-layered architecture,  supporting the creation, execution, monitoring and dynamic update of process instances. 
For efficiency purposes, the interpreter relies on compact bitmap-based encodings of process models.
An experimental evaluation shows that the proposed interpreted approach achieves comparable or lower costs relative to existing compiled approaches.
\end{abstract}

\begin{IEEEkeywords}
Blockchain, Model-Driven Engineering, Business Process Management System, Flexible Process Execution
\end{IEEEkeywords}

\section{Introduction}\label{introduction}

Blockchain technology allows mutually untrusted parties to execute collaborative business processes without relying on a central authority~\cite{Mendling18}. Specifically, blockchain technology allows the parties in a collaborative business process to record the state of the process on a tamper-proof decentralized ledger, which also stores programs (called \emph{smart contracts}) that implement transactions on top of the ledger.

Existing approaches to blockchain-based business process execution are based on the idea of compiling each process model (e.g.\ captured in the Business Process Model and Notation (BPMN)\footnote{https://www.omg.org/spec/BPMN/2.0}) into a set of smart contracts. Once deployed on a blockchain platform, these smart contracts control the instantiation of the process as well as every change to the state of a process instance~\cite{WeberXRGPM16,Mendling18,Lopez-PintadoGDWP18}. In this way, this approach ensures that every process execution abides to its corresponding process model. 
 
In these compiled approaches, the deployed smart contracts are model-dependent. Even a minor change to the process model requires a full re-compilation of the model and the deployment of a new set of smart contracts. Subsequently, new instances of the process are created using the new set of smart contracts, while pre-existing process instances remain tied to the old version of the model. This lack of flexibility is problematic in the context of business processes that are subject to frequent evolution as well as processes with a large number of pathways and exceptions, which typically cannot be captured upfront via a single process model~\cite{ReichertW12}. 
  
 In addition to lack of flexibility, another drawback of these compiled approaches is the inefficiency induced by deployment costs. Indeed, each model (or version thereof) is encoded via separate smart contracts and in contemporary blockchain platforms, such as  Ethereum, every smart contract deployment entails costs proportional to the smart contract's size.
  
 
 
In this setting, this paper addresses the following overarching question: \textit{How to flexibly and cost-efficiently execute collaborative processes involving mutually untrusted parties on a blockchain platform?}
To address this question, the paper puts forward a new approach to blockchain-based business process execution based on an interpreter of BPMN process models. Unlike compiled approaches, the interpreter encodes the semantics of the BPMN language in a single smart contract. As such, the interpreter needs to be deployed only once on the blockchain. The interpreter may be attached to multiple process models, each of which is represented using a dynamically updatable and space-optimized data structure. The interpreter supports the instantiation of any of its associated process models and allows participants to monitor the state of process instances and to execute tasks thereof. A modular architectural design, combined with the use of dynamic data structures, provides flexibility for the participants of the process to react to unexpected situations during the execution. Specifically, the system allows participants to maintain different variants of the same process model or to permanently modify a process model so that all running and future process instances follow the new version of the model. The proposed method has been implemented as an extension of \textit{Caterpillar}~\cite{Lopez-PintadoGDWP18}, which now comprises two process execution engines: a compilation-based and an interpretation-based engine. An experimental evaluation assesses the costs of the interpreted execution approach compared to existing compiled approaches.
 
\begin{figure}[htbp]
\vspace*{-0.5em}
\centerline{\includegraphics[scale=.50]{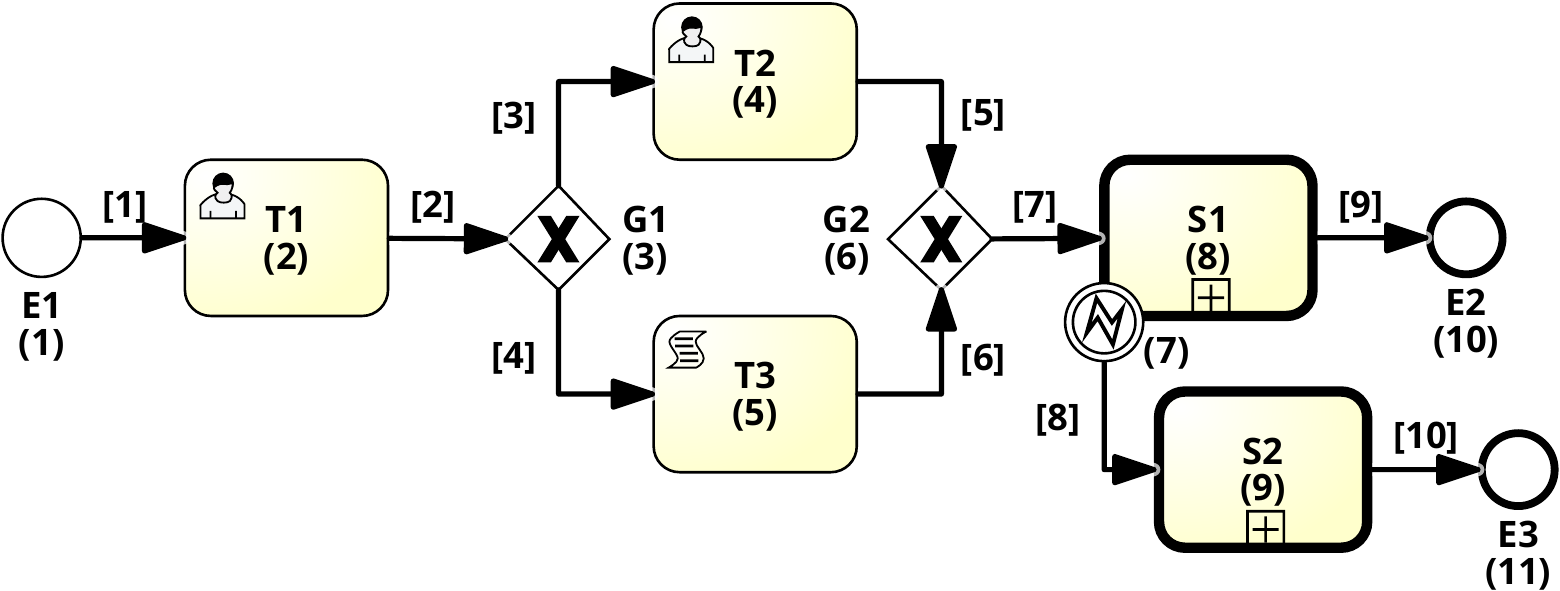}}
\vspace*{-2.0em}
\caption{Simple BPMN model.}
\label{fig:model}
\end{figure}

To illustrate our proposal, we use the BPMN model in Fig.~\ref{fig:model}. Each element is identified by a number. The model contains two user tasks ({\sc T1} and {\sc T2}) to be performed by process participants, and which also serve to check-out/in process data. The remaining elements require no interactions with external actors (they are performed internally). Event {\sc E1} triggers the instantiation of the process, while events {\sc E2} and {\sc E3} end the execution. Gateway {\sc G1} checks conditions, based on the process data, to split the flow into two exclusive paths (joined later via {\sc G2}). Script task {\sc T3} updates the process data by executing internal scripts. Call-activities {\sc S1} and {\sc S2} reference two sub-processes which are modeled separately. An error event may interrupt sub-process {\sc S1}. This error event is caught by the boundary event attached to {\sc S1} and re-directs the flow of control along the exception flow leading to {\sc S2}.
 
The paper is structured as follows. Section \ref{background} discusses blockchain technology and related work. Section \ref{architecture} describes the system's architecture. Section \ref{perspectives} presents the data structures to store the control-flow and data perspectives. Section \ref{interpreter} delves into the inner workings of the interpreter. Finally, Section \ref{experiments} discusses the implementation and evaluation, while Section \ref{conclusion} draws conclusions and sketches future work.

\section{Background and Related Work} \label{background}

 A blockchain is an immutable ledger replicated across a network of untrusted peer nodes~\cite{2019-Blockchain-Book}. The ledger is represented as a linked sequence of blocks. Some nodes, called miners, are responsible for validating and grouping transactions submitted by the users into blocks appended to the blockchain. As no central authority exists, the miners must reach consensus in a distributed manner. In so-called \emph{proof-of-work} blockchains, miners achieve consensus by solving a hard cryptographic puzzle to link a new block to the previous one in the chain. To be accepted, a transaction must be properly formed and cryptographically signed. Miners then send it to different nodes so that the transaction reaches the entire network. 
 
 A smart contract is a program deployed on the blockchain and executed by all the nodes in the network. In the Ethereum blockchain, smart contracts are written in the Solidity language, which is compiled into bytecode and executed on the Ethereum Virtual Machine (EVM). The cost to deploy a contract, which is proportional to its bytecode size, is measured in a unit called \emph{gas}. Once deployed (or instantiated), the smart contract is related to a unique hash address that can be used by external applications to invoke the public functions of the smart contract. Such invocations generate transactions whose cost (also measured in gas) depends on the number and type of the executed instructions~\cite{wood2014ethereum}.
 
 Existing approaches to blockchain-based business process execution compile high-level process models into smart contracts that are deployed and executed on a blockchain platform. For example, \cite{GarciaPDW17, WeberXRGPM16, TranLW18, Nakamura18, Falazi2019} take as input process models specified in BPMN while \cite{madsen2018} compiles models specified in a declarative process modeling language. 
 These approaches, however, suffer from two limitations. First, they focus on the control-flow perspective of process models. In other words, they do not handle process instance data (data perspective) nor the association between resources and tasks (resource perspective).
 One approach that does not suffer from this limitation is our previous proposal, namely Caterpillar~\cite{Lopez-PintadoGDWP18}, which can handle process models with process instance variables, task inputs and outputs, conditional expressions, and task-resource bindings~\cite{Lopez-PintadoDGW19}.
 The second limitation (which is also shared by Caterpillar) is that, due to their reliance on a compilation phase, these approaches do no support the adaptation of a process at runtime, i.e., a change in the model requires the generation and deployment of a new set of smart contracts (which is a costly operation) and existing process instances remain bound to the old version of the model. 
 
 In this paper, we tackle these limitations by taking the basic elements of the Caterpillar approach as a starting point, particularly in regards to handling the data and resource perspectives, but adopting an interpreted approach as opposed to a compiled one. The key idea developed in this paper is to deploy a smart contract, which is able to interpret BPMN models represented by means of a space-optimized data structure that can be modified at any time.  
 
 One of the closest works to ours is presented in~\cite{Sturm18}. This latter proposal also adopts an interpreted approach. However, \cite{Sturm18} focuses on the control-flow perspective (no case variabes) and is limited to a small subset of BPMN elements (tasks and gateways with a simplified execution semantics for join gateways). In particular, it cannot handle subprocesses, error events and boundary events, such as those in Fig.~\ref{fig:model}. Besides, this latter approach bundles together the interpreter with the data structures representing the process model into a single smart contract, and hence suffers from the same flexibility issues as compiled approaches: Any change to the process model requires the deployment of a new smart contract (which is a costly operation) and existing instances remain tied to the previous smart contract. Finally, the approach in \cite{Sturm18} requires that updates to the tasks in the process are performed by a central process owner, which is not suitable in scenarios where there is no central trusted authority.
 
 Another related work that addresses the question of adaptability is~\cite{rybila0HW17}, which proposes an architecture to monitor business processes on the Bitcoin platform. This approach caters for runtime adaptation, but it assumes that the process is not executed on the blockchain itself. The blockchain is used as a recording mechanism (recording the execution of tasks), as opposed to an enforcement mechanism (determining which tasks can be executed) as we do in our approach.
 
 

\section{Architecture} \label{architecture}

The proposed blockchain-based process execution system follows the same design principles as the compiled version of Caterpillar~\cite{Lopez-PintadoGDWP18}. Specifically, the system is designed to enable a set of untrusting parties to develop, deploy and execute collaborative processes on blockchain in a tamper-proof manner. To that end, the full state of the process execution as well as the process execution logic itself are recorded on the blockchain, so that no party is able to execute a transaction that does not abide to the agreed-upon process model. 

\begin{figure*}[htbp]
\centerline{\includegraphics[scale=.50]{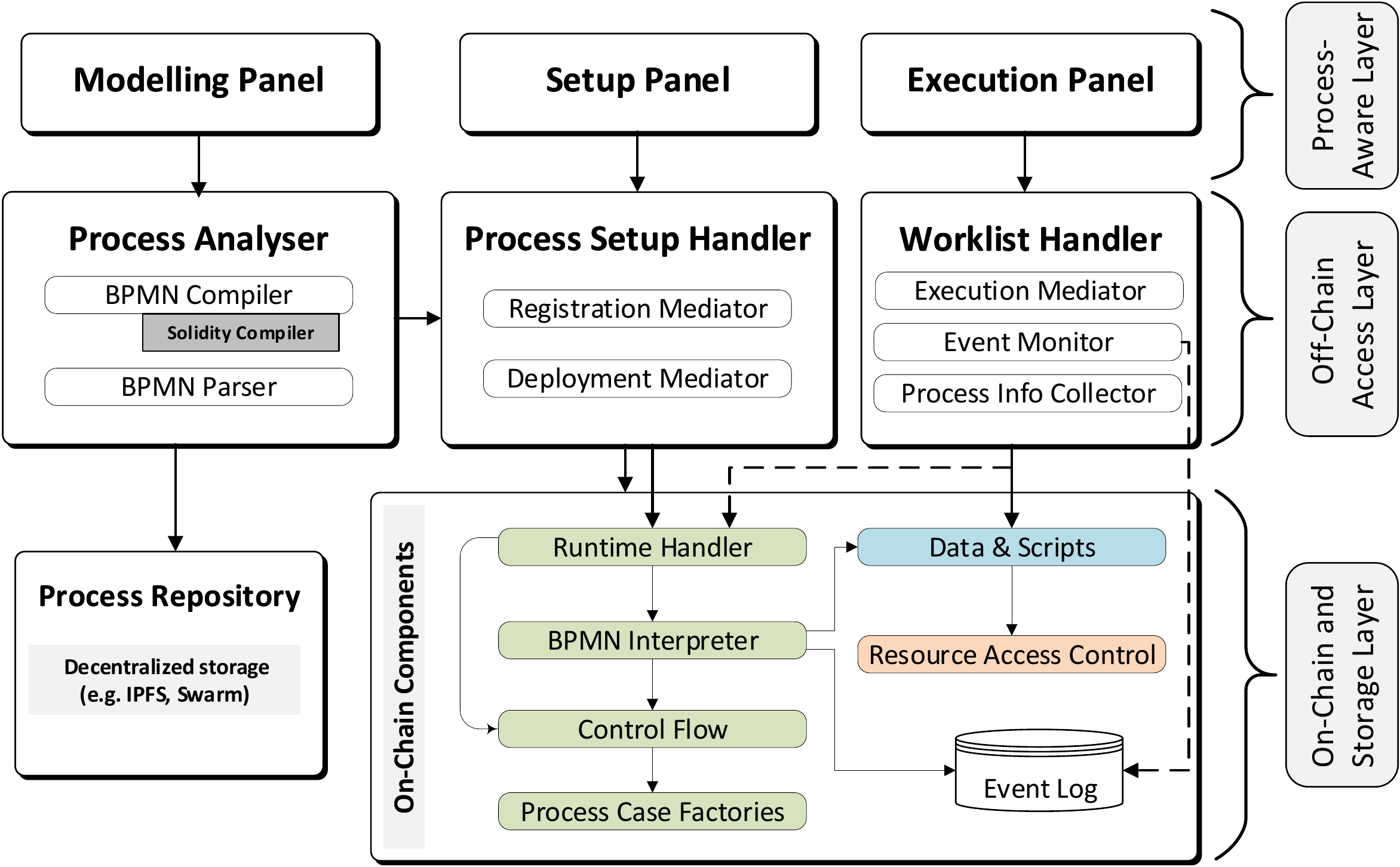}}
\caption{Architecture of the system.}
\label{fig:architecture}
\vspace{-1.0em}
\end{figure*}

Fig. \ref{fig:architecture} illustrates the architecture of the system that is organized into three layers. In the bottom, the {\sc On-Chain and Storage Layer} encloses the set of smart contracts that control the process execution, namely {\sc On-chain Components}, which are replicated across all the nodes of a blockchain network, e.g., the public Ethereum network. Besides, only the data relevant to the process execution is stored {\sc On-chain}, while compilation/parsing artifacts are stored off-chain in the {\sc Process Repository}. In the middle, the {\sc Off-chain Access Layer} includes a set of tools to parse, compile, deploy, execute and monitor business processes in the {\sc On-Chain and Storage Layer}. Finally, the {\sc Process-Aware Layer} comprises a set of components to model/execute the process guided by high level and model-driven interfaces. Note that the components in the {\sc Off-chain Access Layer} and {\sc Process-Aware Layer} run outside of the blockchain, thus they can be tampered with. However, the state of each process instances is stored on-chain, and all the decision points are evaluated on-chain. Besides, each actor in a collaborative process can host the off-chain components separately. Thus, each actor can check directly from the blockchain which actions were performed by others. 

\subsection{On-Chain and Storage Layer}

The {\sc On-Chain Components} are the core that handles the full process execution. In the sake of reusability, the process perspectives, e.g., control flow, data management, and resource allocation, are decoupled into different components. The smart contracts in the components with green background implement the set of common operations to any process model. Thus, they are hard-coded only once based on the BPMN standard and the system requirements. In contrast, the smart contracts in the blue component contain the model-specific data and operations that need to be extracted from each process model. The component with the light orange background handles the user access control and resource allocation. Finally, the event log is provided by some Blockchains, like Ethereum, that gives off-chain components convenient access to update events generated by smart contracts.

The component {\sc Control Flow} stores the information about the structure of the process models, their elements and relations. Given that a process model may include sub-process, the data structure is a tree where each node, named {\sc IFlow}, represents a sub-process that keeps references to its children (if exist). Besides, the nodes map for each enclosed BPMN element the model-related information to be used by the {\sc BPMN Interpreter} to handle the execution. For example, some of the information to store can be the type of element (task, event, gateway, etc.), the incoming/outgoing arcs, to what sub-process an event is attached, and so on. Accordingly, each node needs to be deployed once per sub-process in the model and is identified by the corresponding blockchain address. The address of the root node would identify the full process model. Unlike compiled approaches, the control flow perspective is not statically encoded in a smart contract, but the information is collected off-chain from the model and added to the corresponding nodes dynamically. 

The {\sc Process Case Factory} includes the set of contracts to instantiate and start the execution of a business process. Thus, when a sub-process is linked as a child in the {\sc IFlow} hierarchy, the parent has to store the address of the corresponding factory to instantiate the sub-process during the execution. In the following, we will refer to process instances as process cases, to differentiate them from smart contract instances.

As the name suggests, the smart contracts in the component {\sc Data \& Scripts} implement the data perspective. Data requirements are process-dependent, strongly typed, and their values are conditioned and scoped by the process cases, e.g., a variable defined in a sub-process may have different values in each sub-process case. Accordingly, smart contracts must be compiled from the process model. The scripts related to user/service/script tasks and the conditions to decide the paths in exclusive gateways mostly interact with the process data. Thus, such instructions are also compiled from the model into the contract implementing the data perspective. Besides, the interactions of external actors via user/service tasks also requires two operations per task: check-out to request data from the process case, and check-in to send data to the process and to proceed with the execution. The smart contracts in this component form a hierarchy with a node per (sub-)process case which we will refer to as {\sc IData}. Each {\sc IData} node stores the sub-process state and keeps a reference to the related {\sc IFlow} node. Indeed, the factories linked to a sub-process, in the {\sc IFlow} hierarchy, must define how to instantiate the smart contract of the corresponding {\sc IData} node.

\begin{figure}[htbp]
\centerline{\includegraphics[scale=.42]{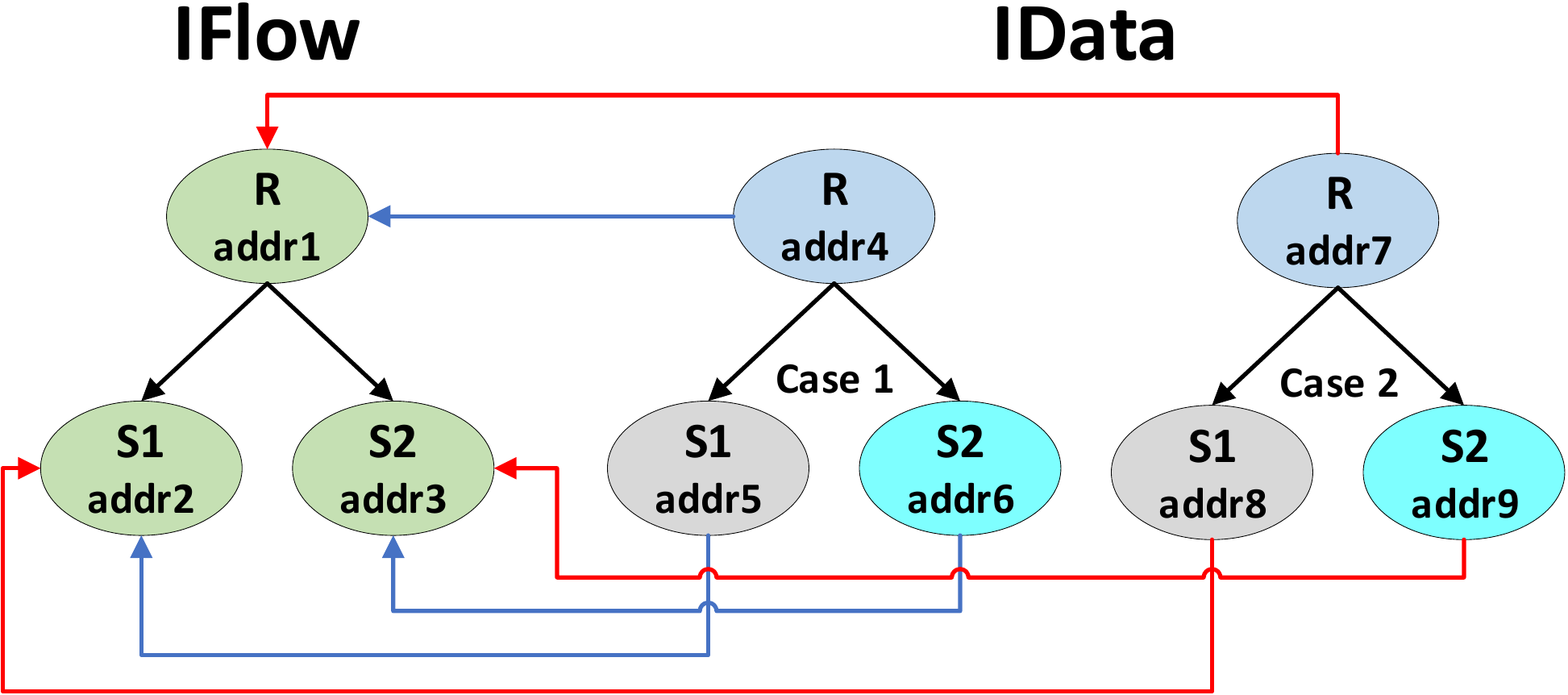}}
\caption{Graphical representation of the the control-flow and data perspectives smart contracts deployed to execute two cases of the process in Fig. \ref{fig:model}.}
\label{fig:hierarchy}
\vspace{-1.5em}
\end{figure}

Fig. \ref{fig:hierarchy} illustrates some of the relations among the smart contracts deployed to execute two cases of the process modeled in Fig. \ref{fig:model}. In Fig. \ref{fig:hierarchy}, each node represents one smart contract instance, while the background colors differentiate the four smart contracts involved. As described above, the control flow perspective uses a single {\sc IFlow} data structure, that is instantiated once per sub-process in the model. On the other hand, three different smarts contracts (one per sub-process) are required on the data perspective to encode common operations of the {\sc IData} structure plus the data/scripts compiled from each sub-process. Each {\sc IData} node keeps a reference to the corresponding {\sc IFlow} node that is used by the {\sc BPMN Interpreter} to check the control flow information of the process case, e.g., to update, after executing an element, which others can be executed. Although not in the figure, only one factory is required per sub-process, e.g., both instances {\tt addr4} and {\tt addr7} of {\tt R} are started by the same factory.

The {\sc Data \& Scripts} component serves as the entry point for external actors to access data and execute tasks. Nevertheless, the access would be restricted by a set of smart contracts in the component {\sc Resource Access Control} derived from binding policies as presented in \cite{Lopez-PintadoDGW19}. Such policies support dynamic bindings of actors into roles, i.e., the process participants reach consensus on who performs which task during the process execution based on binding policies and operations for the nomination, release, and voting. The policies can be extended to restrict not only on the execution of tasks but to also control the operations/updates on the {\sc IFlow} and {\sc IData} data structures as an authorization mechanism for process modification. Such extensions are out of the scope of this paper and left as future work. 

The {\sc BPMN Interpreter} is a single  smart contract that implements the process execution logic defined by the BPMN standard. This component keeps no information about any of the process perspectives, but queries/updates such data from/into the {\sc IFlow} and {\sc IData} structures. 

 The {\sc Runtime Handler} keeps tracks of the process instances, binding policies and their relation with other smart contracts. The {\sc Event Log} provides a source for communication between off-chain and on-chain components. Out of the {\sc On-Chain Components}, the {\sc Process Repository} stores and provides access to compilation artifacts and metadata to link the Solidity code to elements of the BPMN models. The operation of the {\sc Runtime Handler}, {\sc Event Log} and the {\sc Process Repository} is similar to the equivalent components in the architecture of the compiled version of Caterpillar (check \cite{Lopez-PintadoGDWP18} for a complete description).

\subsection{Off-Chain Access and Process-Aware Layers}

The {\sc Off-Chain Access Layer}, in the middle of Fig. \ref{fig:architecture}, provides a service-oriented entry point for external applications to interact with the {\sc On-Chain and Storage Layer}.

The {\sc Process Analyser} (on the left) extends the Caterpillar {\sc BPMN Compiler} to generate the {\sc IData} structure from a BPMN model. The {\sc BPMN Compiler} uses a standard {\sc Solidity Compiler} to produce the metadata and interfaces that are used to deploy and execute the smart contracts. Additionally, the {\sc BPMN Parser} extracts the control flow information from the model that is structured to be inserted in the {\sc IFlow} hierarchy (see Section \ref{perspectives}).

 The {\sc Process Setup Handler} (in the middle) serves as the entry point to deploy smart contracts (e.g., produced by the {\sc BPMN Compiler}), and to update the {\sc IFlow} structure. The {\sc Deployment Mediator} provides the set of operations to deploy the {\sc IFlow} and {\sc IData} hierarchies, as well as the factories and the resource access control contracts. On the other hand, the {\sc Registration Mediator} supports the operations to update the {\sc IFlow} structure (e.g., insert BPMN elements into nodes). Besides, the {\sc Registration Mediator} allows to change the relations among the smart contracts in the {\sc On-Chain Components}, e.g., to link/unlink sub-process as nodes into {\sc IFlow}, update the access control policies, etc.

 On the right of the {\sc Off-Chain Access Layer}, the {\sc Worklist Handler} enables external actors to query the process state and data, as well as to execute tasks on a given process case. The compiled version of Caterpillar implemented the {\sc Worklist Handler} as smart contracts, i.e., on-chain. Besides, the worklist allowed only interaction of human actors, while non-human actors (e.g., information systems, IoT devices) were handled via another on-chain component named {\tt Service Bridge}. In the current approach, we restrict the resource allocation and access control by policies that support dynamic bindings of actors to roles \cite{Lopez-PintadoDGW19} implemented by the {\sc Resource Access Control} component. The latter removes the need to use static worklist contracts generated per process model to validate any data checked-in into the process instances. Thus, the {\sc Worklist handler} is implemented off-chain what is less costly. Besides, the binding policies would verify blockchain accounts that are controlled indistinctly by users, groups, systems, or (IoT) devices. Thus the {\tt Service Bridge} is joined into the {\sc Worklist Handler} off-chain. Accordingly, an actor to check-in a task via the {\sc Execution Mediator} provides the process case (i.e., address of the corresponding {\sc IData} node) and the identifier of the task. Then, the {\sc Execution Mediator} interacts with the {\sc Data \& Scripts} component, that in turn verifies the actor rights via the {\sc Resource Access Control} component. If the actor can perform the task, the process data and state is updated. For that, the corresponding {\sc IData} node invokes the {\sc BPMN Interpreter} that in turn interacts with the related {\sc IFlow} nodes. The {\sc Process Info Collector} interacts with the {\sc Runtime handler} and {\sc Data \& Scripts} to query the active process cases, control flow addresses, the state of a given process case, and to check-out data from user/service tasks. Finally, the {\sc Event Monitor} listens for events generated from the smart contracts, e.g., to notify that the state in some process case was updated.
 
On top of the architecture, the {\sc Process-Aware Layer} exposes the functionality of the {\sc Off-Chain Access Layer} to end users (e.g. process administrators and workers) via a form-based user interface. The rationale of this layer is the same as the {\sc Web Portal} introduced by Caterpillar \cite{Lopez-PintadoGDWP18}. The {\sc Modeling panel} allows the user to draw the BPMN models. The {\sc Setup panel} supports the updates of the data structures of different perspectives and their relations. Finally, the {\sc Execution Panel} interacts with the {\sc Worklist Handler} to retrieve all the information about deployed models, running instances and allow executing tasks by stakeholders.
\section{ Control Flow and Data Representation}\label{perspectives}

As hinted in Section \ref{architecture}, the {\sc IFlow} structure is the tree that captures the hierarchical representation of a process model where each node encloses the control flow information of the corresponding sub-process. During the parsing of the model, each BPMN element is associated with an integer index that is unique per sub-process. Similarly, arcs are numerated as well. The mapping element-index can be accessed from the {\sc Process Repository} that also provides tamper-proof storage. Fig. \ref{fig:model} shows a possible numeration for both arcs and element in the represented process model. Such indexes are used later to encode the elements into bit-sets so the operations can be implemented efficiently using bit-wise operators.

\begin{figure*}[htbp]
\centerline{\includegraphics[scale=.50]{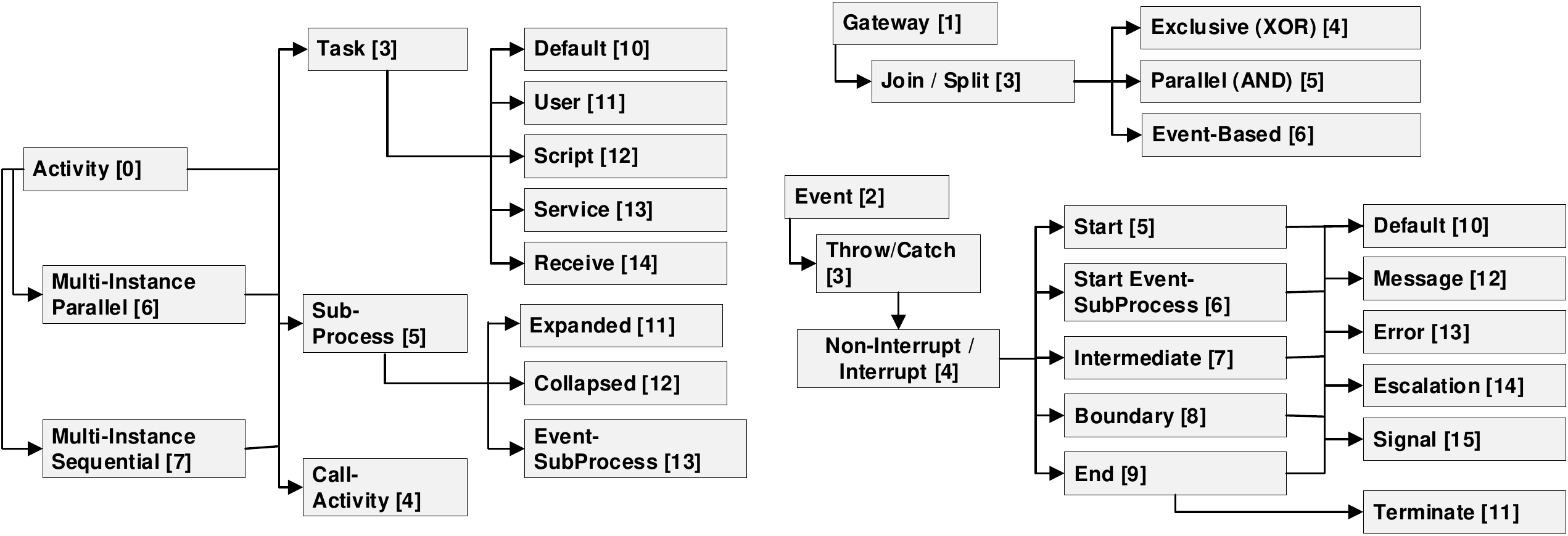}}
\caption{Bit associated to each element/characteristic when encoding the element description as {\tt typeInfo}.}
\label{fig:typeInfo}
\vspace{-1.0em}
\end{figure*}

An {\sc IFlow} node can be updated via three operations. First, the operation {\tt setElement} updates or inserts an element depending on whether it is already contained. The operation requires as input the element index {\tt eInd}, the incoming {\tt preC} and outgoing {\tt postC} arcs, and the element description {\tt typeInfo}. The {\tt preC} and {\tt postC} are bit-sets with 1s in the bits corresponding to the indexes of the arcs contained in the set, and 0s on the remaining. The element description {\tt typeInfo} is also encoded as a bit-set such that each characteristic is identified by a bit (see Fig. \ref{fig:typeInfo}). The second operation {\tt linkSubprocess} add a child into a {\sc IFlow} node. Here, the index of the sub-process/call-activity in the parent must be provided and the address when running the child {\sc IFlow} node. Besides, the number of sub-process instances to create, and the list (can be empty) with the indexes of the attached events are required. Note that the indexes provided must correspond to elements already added in the parent node; otherwise, the operation will be rejected. For example, to link a sub-process to the call-activity labeled as {\sc S1} in Fig. \ref{fig:model}, we have to provide the call-activity index, i.e., 8, the blockchain address of the {\sc IFlow} node created for {\sc S1}, the index of the attached error event, i.e., 7, and one as the number of instances to create. As the factories are related to the {\sc IData} smart contracts, they are updated separately, but before the corresponding element is reached during the execution of a process case. The operations to query the {\sc IFlow} structure are straightforward, thus omitted in the description of this paper.\footnote{The full definition and implementation of the {\sc IFlow} and {\sc IData} can be accessed from \url{ http://git.io/caterpillar}}

Fig. \ref{fig:typeInfo} shows how to encode the element description as {\tt typeInfo} from the bits associated (in brackets) with the elements supported by the Interpreter. For example, user and service tasks are identified by the bits 11 and 13, but as they are also activities and tasks, then they must share those bits (0 and 3) too. Besides, to verify if an element is a user task, the bits 0 and 11 must be checked because the terminate event is also identified by the bit 11, but the bit 2 identifies it as an event. Some bits can encode two characteristics. For example, in the gateways the value 1 in the third bit represents a \textit{join}, otherwise, it is a \textit{split}. Similarly, bits 3 and 4 of an event identify whether it is throwing/catching and interrupting/non-interrupting, respectively.

 Listing \ref{lst:iData} illustrates a sample of the {\sc IData} smart contract generated from the root process in Fig. \ref{fig:model}. To generate the {\sc IData} structure we use the same annotations on the BPMN models defined in \cite{Lopez-PintadoGDWP18}. First, the process variables in lines 3-4 are copied from the global documentation of the model. When compiling a sub-process, such variables are extracted from the documentation of the corresponding element. A single function {\tt execScript} manages the execution of the scripts. It takes as input an element index, executes the scripts associated to the corresponding element, and returns the bit-set with the outgoing arcs to proceed with the process execution, or zero if the element index is not found. Lines 7-16 shows the body of the {\tt execScript} that encodes the exclusive gateway {\sc G1} and the script task {\sc T3}. The outgoing arcs of the exclusive and inclusive gateways contain boolean expressions encoded in Solidity which verifies the process variables. Then based on the evaluation of the expression the execution should be redirected to the corresponding outgoing arc, (cf. lines 8-10). Similarly, the documentation of script tasks includes Solidity instructions to update the process data (cf. lines 11-13).

\begin{lstlisting}[float, belowskip=-2.0 \baselineskip, language=Solidity, caption=Example of the Root IData node produced from model in Fig. \ref{fig:model}.,label={lst:iData},frame=tb]
contract ProcessIData is IData {
   // == PROCESS VARIABLES == 
   bool t1Field;
   bool t2Field;
   // == SCRIPTS TO EXECUTE ==
   function execScript(uint eInd) public returns(uint) {
       if(eInd == 3) { // Gateway G1
           if(t1Field) return 8; // 1 << 3
           else return 16;       // 1 << 4
       } else if(eInd == 5) {    // Script Task T3
           // Execute script defined by the task
           return 64;            // 1 << 6
       }
       return 0;
   }
   // == CHECK IN/OUT FUNCTIONS ==
   function checkIn(uint eInd, bool _input1) public {
       if(eInd == 2)           // User Task T1
           t1Field = _input1;
       else if(eInd == 4)      // User Task T2
           t2Field = _input1;
       revert("Not Found");
   }
   function checkOut(uint eInd) public view returns(bool) {
        if(eInd == 4)           // User Task T2
            return t1Field;
        revert("Not Found");
    }
}
\end{lstlisting}

 The functions {\tt checkIn}, and {\tt chekOut} are generated from the documentation of the elements supporting user interactions in the model. Accordingly, each element should be annotated with expressions of the form $(Data\_To\_Export) : (Data\_To\_Import) -> \{ Operations\_to\_perform \}$, to restrict what data must be read/written from/to the process and what operations must be performed as a result of executing the task. Intuitively, $Data\_To\_Export$ is returned by the corresponding {\tt chekOut} function. On the other hand, the $Data\_To\_Import$ serves as input in the {\tt checkIn} function that also execute $Operations\_to\_perform$. Besides, elements with the same combination of parameter types are grouped into the same function. For example, consider the user tasks {\sc T1} and {\sc T2} in Fig \ref{fig:model} are annotated respectively with:
 $$() : (bool \_t1Field) -> \{t1Field = \_t1Field;\}, $$
 $$(bool t1Field) : (bool \_t2Field) -> \{t2Field = \_t2Field;\}. $$
 As both tasks import a boolean variable, they are encoded in the same {\tt checkIn} function as shown in lines 19-25 of the Listing \ref{lst:iData}. The generation of the {\tt chekOut} function (cf. lines 27-31) only includes {\sc T2} as {\sc T1} contains no data to export. Listing \ref{lst:iData} only illustrates the compiled part of the {\sc IData} nodes. Common operations to access/update the state in process cases and query the structure are are straightforward, thus omitted. 
 
 Each sub-process in the model will produce a {\sc IData} smart contract which is instantiated via a factory that is mapped to the corresponding sub-process in the {\sc IFlow} structure. The hierarchical relationships among the {\sc IData} nodes are built internally during the execution when the corresponding sub-processes are reached in the control flow. Indeed, the {\sc IData} contracts will produce several hierarchy instances, one per process case. External actors are only allowed to create instances of {\sc IData} contracts related to the root process.
 
\section{ BPMN Interpreter Operation } \label{interpreter}

The {\sc BPMN Interpreter} implements six operations to execute process cases based on the BPMN standard. In the following, the notation {\sc IData/IFlow}(address) refers to a node in the corresponding hierarchy (i.e., a smart contract instance) identified by its blockchain address (i.e., variables ending with {\tt Addr}). Besides, the variable {\tt pState} represents the process state, i.e., two bitsets comprising the token distribution on edges and the indexes of sub-processes under execution respectively. For the sake of clarity, the bitwise operations are replaced by functions with names remarked in bold. For example, the functions with suffix {\tt Tokens} would update a bitset representing the edges containing tokens. Further, the keyword {\tt this} is used to invoke the functions implemented in the interpreter. Besides, the types of BPMN elements are written with capital letters. 

Pseudocode in Listing \ref{lst:execute} illustrates the function {\tt executeElements}. It receives as input the blockchain address {\tt iDataAddr} of an {\sc IData} node, and the index {\tt eInd} of the element to be executed. Due to security requirements, external actors do not interact directly with the {\sc BPMN Interpreter}. Instead, an actor checks-in tasks via an {\sc IData} node, that in turn (after verifying the actor privileges) calls {\tt executeElements} to proceed with the process execution. Indeed, lines 2-3 in Listing \ref{lst:execute} would reject any call from addresses distinct to the input {\sc IData} node or the interpreter itself referred as {\tt this}. Elements like gateways, script tasks, throwing events, etc., which not require interaction with external resources, are executed internally.

\begin{lstlisting}[float, belowskip=-2.0 \baselineskip, language=Solidity, caption=Pseudocode of {\tt executeElements} in {\sc BPMN Interpreter}.,label={lst:execute},frame=tb]
function executeElements(iDataAddr, eInd) public
  if (msg.sender != (iDataAddr or this)) 
    throw 'REJECTED'
  iFlowAddr = IData(iDataAddr).getIFlowNode();
  pState = IData(iDataAddr).getSubProcessState();
  queue = new Queue(eInd);
  while(!queue.isEmpty())
    eInd = queue.pop();
    (preC, postC, typeInfo) = IFlow(iFlowAddr).find(eInd);
    if (!isEnabled(preC, typeInfo, pState))
        continue;
    removeTokens(pState, preC);
    switch (typeInfo)
      case PARALLEL_MULTI_INST:
        for(i = 1 to IFlow(iFlowAddr).getCountInst(eInd))
          this.createInst(eInd, iDataAddr);
        addSubProcess(pState, eInd);
      case SEQ_MULT_INST || SUB_PROCESS || CALL_ACTIVITY:
        this.createInst(eInd, iDataAddr);
        addSubProcess(pState, eInd);
      case SCRIPT_TASK || EXCLUSIVE_GATEWAY_SPLIT:
        postC = IData(iDataAddr).executeScript(eInd);
        addTokens(pState, postC);
      case TASKS and GATEWAYS /* Remaining */: 
        addTokens(pState, postC);
      case THROW_EVENT:
        IData(iDataAddr).updateProcessState(pState);
        evtCode = IFlow(iFlowAddr).getEventCode(eInd);
        this.throwEvent(iDataAddr, evtCode, typeInfo);
        pState = IData(iDataAddr).getSubProcessState();
        if(isCompleted(pState))
          return;
        if(INTERMEDIATE_EVENT in typeInfo)
          addTokens(pState, postC);
      default:
        continue;
    foreach (outEInd in IData(iDataAddr).outElements(eInd))
      outInfo = IFlow(iFlowAddr).getTypeInfo(outEInd);
      if(!(EXTERNAL_ELEMENT_INTERACTION in outInfo))
        queue.push(outEInd);
  IData(iDataAddr).updateProcessState(pState);
\end{lstlisting}

Before executing an element, the interpreter requests the related {\sc IFlow} node and the process state from the input {\sc IData} node (see lines 4-5 in Listing \ref{lst:execute}). The candidate elements, starting by the input {\tt eInd}, are added into a queue in the same order they are reached in the control flow. The execution follows a Breadth First Search on the process model until no candidates are available in the queue. In each iteration, the element on the top of the queue is extracted and processed based on its control flow information. First, lines 10-11 check if the element is enabled, i.e., based on the {\tt typeInfo}, checking whether the required tokens are placed on the incoming arcs to enable the element. Such verification uses bitwise operations on the bitsets {\tt postC} and {\tt pState}. If the element is enabled, the tokens on the incoming arcs are removed, i.e., it is not enabled anymore, and the element is executed based on its {\tt typeInfo} (lines 13-36).

Multi-instance activities are split in two cases, both invokes a function {\tt createInstance} implemented by the interpreter (Listing \ref{lst:create}) to create the new {\sc IData} nodes and to update the {\tt pState} with the index of the corresponding sub-process as running (lines 14-20 in Listing \ref{lst:execute}). Parallel multi-instances produces as many nodes as specified by the model. On the other hand, sequential multi-instances generate only the first node, because the other nodes require the completion of the process case represented by the previous node, which involves the catching of an end event. Thus, they are instantiated by another function called {\tt tryCatchEvent} (See Listing \ref{lst:catch}). 

Script tasks and exclusive split gateways in lines 21-23 of Listing \ref{lst:execute} executes the corresponding script compiled from the process model into the {\sc IData} node, which also returns the outgoing arc to update the process state. The remaining tasks and gateways, not involving scripts, are executed by adding the tokens in {\tt postC} to the process state as shown in lines 24-25. Note that tasks checked-in by external actors may also include scripts, but they are executed by the corresponding check-in function in the {\sc IData} node. Accordingly, only the external task received as input in the {\tt executeElements} is added into the queue, i.e., they are never added/executed internally by the interpreter during the execution (see line 40). 

Continuing with in Listing \ref{lst:execute}, lines 26-34 handle the throwing of events by calling the function {\tt throwEvent}. The propagation of the event across the {\sc IData} hierarchy may provoke the interruption of the sub-process represented by the corresponding node, e.g., as a result of handling an error event. Thus, the execution continues only if the process state contains some element enabled after the propagation (lines 31-32). In the case of intermediate throwing events, their outgoing arcs must be added to the process state (lines 33-34). Finally, the loop in lines 37-40 adds each adjacent element (reached via an outgoing arc) as a candidate into the execution queue.

Listing \ref{lst:create} illustrates the sequence of steps required to create a node in the {\sc IData} hierarchy in a given process case. The input is the {\sc IData} node to be the parent and the index associated the child sub-process in the {\sc IFlow} structure. First, the factory mapped to the corresponding sub-process is requested, throwing an error if no factory exists (lines 3-6). Next, the new child {\sc IData } node is created via the factory, and the relation parent-child is updated (lines 7-9). Finally, the function {\tt executeElements} is performed in the new child, to ensure that only elements that require external interaction remain enabled. A particular case occurs when an external actor instantiates a root node. There, the root {\sc IFlow} node must be provided, and no relation parent-child is added. 

\begin{lstlisting}[float, belowskip=-2.0 \baselineskip, language=Solidity, caption=Pseudocode of {\tt createInstance} in {\sc BPMN Interpreter}.,label={lst:create},frame=tb]
function createInstance(iDataAddr, eInd)
  iFlowAddr = IData(iDataAddr).getIFlowNode();
  chIFlowAddr = IFlow(iFlowAddr).getChildIFlow(eInd);
  factoryAddr = IFlow(chIFlowAddr).getFactory();
  if(factoryAddr == address(0)) 
    throw 'REJECTED' 
  chIDataAddr = IFactory(factoryAddr).newInstance();
  IData(chIDataAddr).setParent(iDataAddr, chIFlowAddr);
  IData(iDataAddr).addChild(eInd, chIDataAddr);
  eInd = IFlow(chIFlowAddr).getInitElement();
  this.executeElements(chIDataAddr, eInd);
\end{lstlisting}

Listing \ref{lst:throw} illustrates some validations to perform before propagating a thrown event to the parent. The input is the {\sc Idata} where the event is thrown, the event code (required for catching purposes) and its {\tt typeInfo}. First, if the event is a message, a blockchain event will be written in the event log to notify that a certain point of the process execution was reached (lines 3-4). Default and Message end events are propagated to the parent only if the execution in the current node (sub-process) is finished (lines 5-7). Remaining events, i.e., error, escalation, terminate, always propagate to the parent. The terminate event must stop the current node before the propagation performed by the function {\tt killSubProcess}.

\begin{lstlisting}[float, belowskip=-2.0 \baselineskip, language=Solidity, caption=Pseudocode of {\tt throwEvent} in {\sc BPMN Interpreter}.,label={lst:throw},frame=tb]
function throwEvent(iDataAddr, evtCode, typeInfo)
  pState = IData(iDataAddr).getSubProcessState();
  if(MESSAGE in typeInfo) 
    emit MessageSent(evtCode);
  if(DEFAULT or MESSAGE in typeInfo)
    if(isCompleted(pState))
      this.tryCatchEvent(iDataAddr, evtCode, typeInfo);
  else
    if(TERMINATE in typeInfo))
       this.killSubProcess(iDataAddr);
    this.tryCatchEvent(iDataAddr, evtCode, typeInfo);
\end{lstlisting}

Listing \ref{lst:catch} describes how an event thrown from a node is handled in the parent. The input is the {\sc IData} node where the event was thrown, the event code and its {\tt typeInfo}. First, the propagation is stopped in lines 3-6 if no parent exists, and finishing the process execution if the received event is an error. Lines 7-10 queries the information about parent and child stored into variables with prefixes `catch' and `subP' respectively. Then, if the execution in the child is completed, the parent updates the state by removing the sub-process and adding a token on its outgoing arc (lines 11-17). In the case of sequential multi-instance activities, lines 18-19, if any instance is pending to be created, then the next {\sc IData} node is created (cf. first node is created by function {\tt executeElements}). 

\begin{lstlisting}[float, belowskip=-1.5 \baselineskip, language=Solidity, caption=Pseudocode of {\tt tryCatchEvent} in {\sc BPMN Interpreter}.,label={lst:catch},frame=tb]
function tryCatchEvent(iDataAddr, evtCode, typeInfo)
  catchIDataAddr = IData(iDataAddr).getParent();
  if(catchIDataAddr == address(0))
    if(ERROR in typeInfo)
      this.killSubProcess(iDataAddr);
    return;
  catchIFlowAddr = IData(catchIDataAddr).getIFlowNode();
  pState = IData(iDataAddr).getSubProcessState();
  subPInd = IData(iDataAddr).getIndexInParent();
  subPInfo =  IFlow(catchIFlowAddr).getTypeInfo(subPInd);
  if(isCompleted(IData(iDataAddr).getSubProcessState()))
    IData(catchIDataAddr).decreaseInstCount(subPInd);
  subPCount = IData(catchIDataAddr).getCountInst(subPInd);
  if(subPCount == 0)
    removeSubProcess(catchIDataAddr, subPInd);
    postC = IFlow(catchIFlowAddr).getPostC(subPInd);
    addTokens(catchIDataAddr, postC);
  else if(SEQ_MULTI_INST in subPInfo)
      this.createInstance(catchIDataAddr, subPInd);
  if(!(MESSAGE or DEFAULT or TERMINATE) in typeInfo)
    if(SIGNAL in typeInfo)
      while(catchIDataAddr != address(0))
        iDataAddr = catchIDataAddr;
        catchIDataAddr = IData(iDataAddr).getParent();
      IData(iDataAddr).updateProcessState(pState);
      this.broadcastSignal(iDataAddr);
      return;
    foreach(ev in IFlow(catchIFlowAddr).getEventList())
      if(IFlow(catchIFlowAddr).getEvtCode(ev) == evtCode)
        evInfo = IFlow(catchIFlowAddr).getTypeInfo(ev);
        attchTo = IFlow(catchIFlowAddr).getAttachedTo(ev);
        if(EVENT_SUB_PROCESS_START in evInfo)
          if(INTERRUPTING in evInfo)
            this.killSubProcess(catchIDataAddr);
            pState = EMPTY;
          this.createInstance(catchIDataAddr, attchTo);
          addSubProcess(pState, attchTo);
          IData(catchIDataAddr).updateProcessState(pState);
          return;
        else if(BOUNDARY in evInfo && attchTo == subPInd)
          if(INTERRUPTING in evInfo)
            this.killSubProcess(iDataAddr);
            removeSubProcess(pState, subPInd);
          postC = IFlow(catchIFlowAddr).getPostC(ev);
          addTokens(pState, posC);
          IData(catchIDataAddr).updateProcessState(pState);
          next = IFlow(catchIFlowAddr).getOutElement(ev);
          this.executeElements(catchIDataAddr, next);
          return;
    this.throwEvent(catchIDataAddr, evtCode, typeInfo)
\end{lstlisting}

Events like terminate, message and default are propagated to the parent to notify that a child is finished. Thus they are not caught by another event as for signals, errors, and escalations. Lines 21-26 in Listing \ref{lst:catch} shows how signal events are propagated to the root, and later broadcast to each running sub-process. Errors and escalations are handled by the parent if exist a catching event matching the code of the event propagated (lines 28-29). If event is matched, two cases may occur. First in lines 32-39, if the event is caught to start an event sub-process, then the parent has to be killed if the catching event was modeled as interrupting, and a new instance of the event-sub-process is created as the only child. Otherwise, the event-sub-process runs in parallel with the enabled elements in the parent. The second case occurs if the event is caught in the boundary of the sub-process that is throwing it (lines 40-49). Then, the sub-process is ended if the catching event is marked as interrupting. Also, a token is added on the outgoing arc of the boundary event, and the execution proceeds by calling the function {\tt executeElements}. Finally, in line 50 the event is re-thrown if it cannot be caught in the current node. 

The remaining two functions, {\tt killElements} and {\tt broadcastSignal}, traverse each descendant reachable from a source node. The function {\tt killElement} takes a {\sc IData} node as input and updates the process state in such node as empty (all bits are set to 0), repeating recursively the same procedure for each child that remains running. The function {\tt broadastSignal} performs a strategy similar as in lines 28-49 of Listing \ref{lst:catch}, but catching only signal events, for each running sub-process from the root {\sc IData} node. As the only difference, signals do not require to match the {\tt evtCode}, i.e., each catching signal attached/contained to/in an enabled sub-process must be handled.

\section{ Implementation and Evaluation } \label{experiments}

The {\sc Off-Chain Access Layer} and {\sc Process-Aware Layer} are implemented in {\tt Node.js}\footnote{https://nodejs.org/en/}. The smart contracts are compiled using the standard Solidity compiler {\tt solc-js}\footnote{https://github.com/ethereum/solc-js}. The deployment and interaction with running instances of the smart contracts are supported via the Ethereum client {\tt Geth}\footnote{https://github.com/ethereum/go-ethereum/wiki/geth}. The full implementation of the system proposed in this paper can be downloaded under the BSD 3-clause “New” or “Revised” License from the Caterpillar's repository at \url{https://github.com/orlenyslp/Caterpillar}, version V3.0.

\begin{table*}[htbp]
\caption{Caterpillar Interpreter REST API.}
\vspace{-0.7em}
\label{table:restAPI}
\centering
\begin{tabular}{|l|l|p{9cm}|}
\hline
\textbf{Verb} & \textbf{URI} & \textbf{Description} \\
\hline
POST & /interpreter & Creates a new instance of the {\sc BPMN Interpreter} \\
\hline
POST & /interpreter/models & Parses a BPMN model. This operation may update the required {\sc On-Chain Components} and {\sc Process Repository} (if specified), thus the process would be ready to be executed. \\
\hline
GET & /interpreter/models/ & Retrieves the list of parsed BPMN models \\
\hline
GET & /interpreter/models/:m-hash & Retrieves a BPMN model, its compilation artifacts and {\sc IFlow} root node instances \\
\hline
POST & /i-flow & Creates a empty {\sc IFlow} node  \\
\hline
PATCH & /i-flow/element/:cf-address & Updates a BPMN element into a given {\sc IFlow} node  \\
\hline
PATCH & /i-flow/child/:cf-address & Links a child node (i.e. associated to a sub-process) in a given {\sc IFlow} node  \\
\hline
PATCH & /i-flow/factory/:cf-address & Relates a factory with a sub-process in a given {\sc IFlow} node (i.e. a related {\sc IData} smart contract must exist)  \\
\hline
GET & /i-flow/:cf-address/ & Retrieves the information (i.e., elements, child sub-process and factories addresses) from a given {\sc IFlow} node  \\
\hline
POST & /i-flow/p-cases/:cf-address & Creates a new process case from a given {\sc IFlow} root node. \\
\hline
GET & /i-flow/p-cases/:cf-address & Retrieves all the process cases created (i.e. {\sc IData} instances) from a given {\sc IFlow} root node. \\
\hline
GET & /i-data/:pc-address & Retrieves the current state of a given process case \\
\hline
GET & /i-data/:pc-address/i-flow/:e-index & Checks-out a task in a given process case  \\
\hline
PATCH & /i-data/:pc-address/i-flow/:e-index & Checks-in a task in a given process case \\
\hline
\end{tabular}
\vspace{-1.0em}
\end{table*}

The functionality of the {\sc Process-Aware Layer} is exposed via the REST API described in Table \ref{table:restAPI}. The REST API is built around three types of resources: (i) {\tt interpreter} which manages the deployment of the {\sc BPMN Interpreter} and the operations derived from the parsing of the models, (ii) {\tt i-flow} which involves the deployment and interactions with {\sc IFlow} nodes, e.g., to update BPMN elements, link sub-processes and factories, create new process instances, etc. and (iii) {\tt i-data} which refers to the interactions with {\sc IData} nodes, e.g., to verify the process state, and check-in/out tasks. The full documentation of the REST API, including the format of the messages used on the requests/responses of each operation, can be found in the Caterpillar's repository.

In the following, we describe an experimental evaluation aimed at assessing the costs of executing business processes using the interpreted approach presented in this paper, relative to existing compiled solutions on blockchain-based process execution~\cite{WeberXRGPM16, GarciaPDW17, Lopez-PintadoGDWP18}. Accordingly, we used the same four datasets, consisting of a BPMN model and the corresponding event log. The first dataset, named \textit{Invoicing}, corresponds to a real-world business process, used and distributed by Minit\footnote{http://www.minitlabs.com/}. The other datasets referred to as \textit{Supply chain}, \textit{Incident mgmt.} and \textit{Insurance claim} were extracted from the literature and used on the experiments reported in~\cite{WeberXRGPM16}. 

Like in the compiled version of Caterpillar~\cite{Lopez-PintadoGDWP18}, we implemented a component which replays the distinct log traces interacting with the REST API described in Table \ref{table:restAPI}. The replayer parses each of the four BPMN models in the datasets, deploys the contracts of the {\sc BPMN Interpreter} and {\sc IFlow} nodes, and updates each {\sc IFlow} node with the corresponding BPMN elements and factories. Once the configuration of the models is completed, the replayer reads the corresponding log, and sequentially instantiates of each process case ({\sc IData} node) and executes the corresponding events in the log via the REST API. Besides, the replayer collects and assesses the gas consumed by each operation once the corresponding transaction is included in the blockchain. The experiments were performed on a {\tt Node.js} based Ethereum client named ganache-cli\footnote{https://github.com/trufflesuite/ganache-cli} which simulates a full client for developing and testing purposes on Ethereum.

Table \ref{tbl:config} presents the costs in gas derived from setting-up the {\sc IFlow} structure at runtime (not required by the compiled approaches). The column labeled as \textit{Avg. Reg. Cost} shows the average costs of registering a BPMN element into the corresponding {\sc IFlow} node. Besides, the deployment of the interpreter costs 3,365,098 gas, while deploying a single {\sc IFlow} node costs 721,049 gas.

\begin{table}[htbp]
\vspace{-1.2em}
\caption{Registration Costs of BPMN elements.}
\vspace{-1.2em}
\label{tbl:config}
\center
\begin{tabular}{|l|c|c|}
\hline
\bf Process & \bf BPMN Elements & \bf Avg. Reg. Cost.  \\
\hline
Invoicing & 60 & 110,760 \\
\hline
Supply chain & 15 & 105,516 \\
\hline
Incident mgmt. & 18 & 114,671 \\
\hline
Insurance claim & 24 & 112,850 \\
\hline
\end{tabular}
\vspace{-0.7em}
\end{table}

Table \ref{tbl:results} shows the gas consumption observed in the experiments. For comparison, we used three baselines, in addition to the approach described by this paper (labeled as \textit{I- Caterp} in Table \ref{tbl:results}). The first baseline (labeled \textit{Default}) corresponds to the approach presented in~\cite{WeberXRGPM16} which compiles the control-flow perspective into a smart contract that also stores one boolean variable per (binary) decision gateway in order to determine which conditional flow should be selected. The second baseline (labeled \textit{Opt- CF}) is similar to \textit{Default} but it uses reduction rules to simplify the control-flow structure of the process model prior to compilation~\cite{GarciaPDW17}.\footnote{In the \textit{Default} and \textit{Opt- CF} approaches, one smart contract is deployed per process instance. In~\cite{GarciaPDW17}, a second optimized approach (\textit{Opt-Full}) is proposed wherein all instances of a process are executed by a single smart contract, thus leading to lower instantiation costs. However, this approach cannot be extended to deal with data and resources because, when data is involved, one contract per process instance is needed to hold the instance data. Given this fundamental limitation, we exclude this approach from this comparison.} These baselines focus on the control-flow perspective. They do not handle the data and resource perspective (i.e. storing data attributes and managing work-items). The fourth baseline, named \textit{C- Caterp} corresponds to the compiled version of Caterpillar, proposed in~\cite{Lopez-PintadoGDWP18}, which provides a more advanced architecture, capable of handling the data and resource perspectives. In all the cases, Table \ref{tbl:results} shows the average costs to instantiate the processes and to execute a trace in the event log.

\begin{table}[htbp]
\vspace{-1.2em}
\caption{Process Instantiation and Execution Costs.}
\vspace{-1.5em}
\label{tbl:results}
\center
\newcommand{\mywidth}{3.5em}
\begin{tabular}{|p{3.5em}|c|l|r|r|}
\hline
{\bf Process} & \parbox[t][1pt][t]{0.7cm}{\bf Traces} & \parbox[t][10pt][t]{1.20cm}{\bf Approach} & \multicolumn{2}{c|}{\bf Average Cost} \\
\cline{4-5}
&&&{\bf Instant.} & {\bf Exec.} \\
\hline
\multirow{4}{\mywidth}{Invoicing} & \multirow{4}{*}{5316} & Default & 1,089,000 & 383,109 \\
\cline{3-5}
&& Opt- CF & 807,123 & 297,351 \\
\cline{3-5}
&& \bf C- Caterp & \bf 2,830,063 & \bf 1,088,315 \\
\cline{3-5}
&& \bf I- Caterp & \bf 543,503 & \bf 652,784 \\
\hline
\multirow{4}{\mywidth}{Supply chain} & \multirow{4}{*}{62} & Default & 304,084 & 281,206 \\
\cline{3-5}
&& Opt- CF & 298,564 & 272,186 \\
\cline{3-5}
&& \bf C- Caterp &  \bf 1,100,590 & \bf 566,861 \\
\cline{3-5}
&& \bf I- Caterp &  \bf 434,891 & \bf 418,259 \\
\hline
\multirow{4}{\mywidth}{Incident mgmt.} & \multirow{4}{*}{124} & Default & 365,207 & 185,680 \\
\cline{3-5}
&& Opt- CF & 345,743 & 166,345 \\
\cline{3-5}
&& \bf C- Caterp & \bf 1,119,803 & \bf 324,420 \\
\cline{3-5}
&& \bf I- Caterp & \bf 496,038 & \bf 273,811 \\
\hline
\multirow{4}{\mywidth}{Insurance claim} & \multirow{4}{*}{279} 
& Default & 439,143 & 552,274 \\
\cline{3-5}
&& Opt- CF & 391,510 & 514,712 \\
\cline{3-5}
&& \bf C- Caterp & \bf 1,338,152 & \bf 1,235,617 \\
\cline{3-5}
&& \bf I- Caterp & \bf 500,614 & \bf 992,461 \\
\hline
\end{tabular}
\end{table}

Table \ref{tbl:results} shows that, in all the cases, the interpreter consumes significantly less gas than the compiled version of Caterpillar. This result was expected given that: (1) the deployment costs are amortized as the number of process instances grows, given that the interpreter and the {\sc IFlow} structure are reused, and (2) redundancies in the code generation, present in compiled approaches, are eliminated, resulting in the reduction of the size of the smart contracts. Table \ref{tbl:results} also shows the costs of the interpreter are relatively close to the approaches represented by \textit{Default} and \textit{Opt- CF}. Although in most of the cases the interpreter consumed more gas, it is worth noting that the comparison is not straightforward as \textit{Default} and \textit{Opt- CF} mainly focus on the control-flow perspective. In contrast, the interpreter implements a more advanced and flexible architecture which handles the three process perspectives, and also more advanced control-flow elements like sub-processes, multi-instances, and event propagation.

\section{ Conclusion }\label{conclusion}

This paper presented a blockchain-based execution engine for collaborative business processes. Unlike previous approaches that rely on compilation of BPMN models into smart contracts, the proposed engine relies on a BPMN interpreter that takes as input a space-optimized representation of process models. This design reduces the costs of deployment since the smart contract encoding the interpreter only needs to be deployed once. 
It also allows participants to make changes to the process model in a way that these changes can be applied both to new process instances and to already running instances.

The proposal has been implemented in the Caterpillar blockchain-based process execution system, in such a way that Caterpillar now supports both a compiled and an interpreted execution approach. 
The empirical evaluation shows that the interpreted approach is more cost-efficient than the compiled one. In addition, despite supporting all three process modeling perspectives (control-flow, data, and resources), the costs of the Caterpillar interpreter are comparable to those of existing baselines that only support the control-flow perspective.

The fact that the proposed approach allows participants to dynamically update a process model raises the question of how to ensure that the already running instances do not end up in an inconsistent state after a process model change. For example, replacing a pair of XOR gateways with AND gateways may put some instances in an inconsistent state, possibly leading to a deadlock. A direction for future work is to adapt existing approaches for consistency verification of dynamic process model changes to this setting~\cite{FlexibleBook}. Another avenue of future work is to extend the approach with policies that restrict the allowed changes and/or that allow participants to selectively accept or reject changes at runtime.

\bibliographystyle{IEEEtran}  
\bibliography{bibliography}

\begin{thebibliography}{10}
\providecommand{\url}[1]{#1}
\csname url@samestyle\endcsname
\providecommand{\newblock}{\relax}
\providecommand{\bibinfo}[2]{#2}
\providecommand{\BIBentrySTDinterwordspacing}{\spaceskip=0pt\relax}
\providecommand{\BIBentryALTinterwordstretchfactor}{4}
\providecommand{\BIBentryALTinterwordspacing}{\spaceskip=\fontdimen2\font plus
\BIBentryALTinterwordstretchfactor\fontdimen3\font minus
  \fontdimen4\font\relax}
\providecommand{\BIBforeignlanguage}[2]{{%
\expandafter\ifx\csname l@#1\endcsname\relax
\typeout{** WARNING: IEEEtran.bst: No hyphenation pattern has been}%
\typeout{** loaded for the language `#1'. Using the pattern for}%
\typeout{** the default language instead.}%
\else
\language=\csname l@#1\endcsname
\fi
#2}}
\providecommand{\BIBdecl}{\relax}
\BIBdecl

\bibitem{Mendling18}
J.~Mendling and et~al., ``Blockchains for business process management -
  challenges and opportunities,'' \emph{{ACM} Trans. Management Inf. Syst.},
  pp. 4:1--4:16, 2018.

\bibitem{WeberXRGPM16}
I.~Weber and et~al., ``{Untrusted Business Process Monitoring and Execution
  Using Blockchain},'' in \emph{{BPM}}, 2016, pp. 329--347.

\bibitem{Lopez-PintadoGDWP18}
O.~L{\'{o}}pez{-}Pintado, L.~Garc{\'{\i}}a{-}Ba{\~{n}}uelos, M.~Dumas,
  I.~Weber, and A.~Ponomarev, ``{Caterpillar:} {A} business process execution
  engine on the {Ethereum} blockchain,'' \emph{Software: Practice and
  Experience}, 2019, in press, accepted April 2019.

\bibitem{ReichertW12}
M.~Reichert and B.~Weber, \emph{Enabling Flexibility in Process-Aware
  Information Systems - Challenges, Methods, Technologies}.\hskip 1em plus
  0.5em minus 0.4em\relax Springer, 2012.

\bibitem{2019-Blockchain-Book}
X.~Xu, I.~Weber, and M.~Staples, \emph{Architecture for blockchain
  applications}.\hskip 1em plus 0.5em minus 0.4em\relax Springer, 2019.

\bibitem{wood2014ethereum}
G.~Wood, ``Ethereum: A secure decentralised generalised transaction ledger,''
  2017.

\bibitem{GarciaPDW17}
L.~Garc{\'{\i}}a{-}Ba{\~{n}}uelos, A.~Ponomarev, M.~Dumas, and I.~Weber,
  ``Optimized execution of business processes on blockchain,'' in \emph{{BPM}},
  2017, pp. 130--146.

\bibitem{TranLW18}
A.~Tran, Q.~Lu, and I.~Weber, ``Lorikeet: {A} model-driven engineering tool for
  blockchain-based business process execution and asset management,'' in
  \emph{Demo Track at {BPM}}, 2018, pp. 56--60.

\bibitem{Nakamura18}
H.~Nakamura, K.~Miyamoto, and M.~Kudo, ``Inter-organizational business
  processes managed by blockchain,'' in \emph{WISE}, 2018, pp. 3--17.

\bibitem{Falazi2019}
G.~Falazi, M.~Hahn, U.~Breitenb{\"u}cher, and F.~Leymann, ``Modeling and
  execution of blockchain-aware business processes,'' \emph{SICS
  Software-Intensive Cyber-Physical Systems}, 2019.

\bibitem{madsen2018}
M.~Madsen and et~al., ``Collaboration among adversaries: distributed workflow
  execution on a blockchain,'' in \emph{Symposium on Foundations and
  Applications of Blockchain}, 2018.

\bibitem{Lopez-PintadoDGW19}
O.~L{\'{o}}pez{-}Pintado, M.~Dumas, L.~Garc{\'{\i}}a{-}Ba{\~{n}}uelos, and
  I.~Weber, ``Dynamic role binding in blockchain-based collaborative business
  processes,'' in \emph{{CAiSE}}, 2019, in press, accepted February 2019.

\bibitem{Sturm18}
C.~Sturm, J.~Szalanczi, S.~Sch{\"{o}}nig, and S.~Jablonski, ``A lean
  architecture for blockchain based decentralized process execution,'' in
  \emph{Workshops at {BPM}}, 2018, pp. 361--373.

\bibitem{rybila0HW17}
C.~Prybila, S.~Schulte, C.~Hochreiner, and I.~Weber, ``Runtime verification for
  business processes utilizing the bitcoin blockchain,'' \emph{Future
  Generation Computer Systems}, 2017.

\bibitem{FlexibleBook}
M.~Reichert and B.~Weber, \emph{Enabling Flexibility in Process-Aware
  Information Systems - Challenges, Methods, Technologies}.\hskip 1em plus
  0.5em minus 0.4em\relax Springer, 2012.

\end{thebibliography}

\end{document}